# THE RELATIONSHIP OF TIME AND SPACE

S. B. M. Bell and B. M. Diaz

Interdisciplinary Quantum Group, Department of Computer Science

The University of Liverpool, Chadwick Building, Peach Street

Liverpool, L69 7ZF, United Kingdom

Email: Mail@SarahBell.org.uk and B.M.Diaz@csc.liv.ac.uk

## ABSTRACT

We show that, in addition to radiation travelling at the speed of light, QED theory predicts a second type of radiation with an infinite velocity. We also show that charge, as it appears in the Dirac equation, may have a triune nature.

## 1. PRELIMINARIES

### 1.1 Note on Nomenclature

Matrices consisting of more than one row or column, quaternions, maps, and lifts are given boldface type. * signifies complex conjugation. $^T$ signifies transposition. $^\dagger$ signifies Hermitian conjugation. $^\ddagger$ signifies quaternion conjugation [Altmann 1986]. A lowercase Latin subscript stands for 1, 2 or 3 and indicates the space axes. A lowercase Greek subscript stands for 0, 1, 2, or 3 and indicates the spacetime axes. $i$ is the square root $-1.\,\mathbf{i}_0 = 1.\,\mathbf{i}_1 = \mathbf{i},\ \mathbf{i}_2 = \mathbf{j}, \mathbf{i}_3 = \mathbf{k}$ stand for the quaternion matrices, where $\mathbf{i}_r^2 = -1$ and $\mathbf{i}_1\mathbf{i}_2 = \mathbf{i}_3,\ \mathbf{i}_2\mathbf{i}_1 = -\mathbf{i}_3$ with cyclic variations [Altmann 1986]. We use $-i\boldsymbol{\sigma}_r$ as the basis, where $\boldsymbol{\sigma}_r$ are the Pauli matrices [Dirac 1958],

$$\boldsymbol{\sigma}_1 = \begin{pmatrix} 0 & 1 \\ 1 & 0 \end{pmatrix},\ \boldsymbol{\sigma}_2 = \begin{pmatrix} 0 & -i \\ i & 0 \end{pmatrix},\ \boldsymbol{\sigma}_3 = \begin{pmatrix} 1 & 0 \\ 0 & -1 \end{pmatrix}.$$

Since $i^\ddagger = i \neq i^\dagger = -i,$ this means that $\mathbf{i}_1{*} = -\mathbf{i}_1, \mathbf{i}_2{*} = \mathbf{i}_2, \mathbf{i}_3{*} = -\mathbf{i}_3, \mathbf{i}_1^T = \mathbf{i}_1,$ $\mathbf{i}_2^T = -\mathbf{i}_2, \mathbf{i}_3^T = \mathbf{i}_3, \mathbf{i}_r^\dagger = -\mathbf{i}_r = \mathbf{i}_r^\ddagger.$ We define the underline symbol and the vertical bar as indicating *a reflector* or *a rotator* matrix, respectively [Bell et al. 2000],

$$\underline{\mathbf{U}} = \underline{\mathbf{U}}(\mathbf{Q},\mathbf{U}) = \begin{pmatrix} 0 & \mathbf{Q} \\ \mathbf{U} & 0 \end{pmatrix},\ \mathbf{U}| = \mathbf{U}|\,(\mathbf{Q},\mathbf{U}) = \begin{pmatrix} \mathbf{Q} & 0 \\ 0 & \mathbf{U} \end{pmatrix}$$

where $\mathbf{Q}$ and $\mathbf{U}$ are quaternions. $\underline{^T}$ signifies *reflector transposition*, $\underline{\mathbf{U}}^{\underline{T}}(\mathbf{Q},\mathbf{U}) = \underline{\mathbf{Q}}(\mathbf{U},\mathbf{Q}).$ We use units in which $c = h/2\pi = 1$ where $c$ is the speed of light and $h$ is Planck's constant and may also stand for a gravitational version [Bell and Diaz 2004], [Bell 2004]. Repetition of a subscript does not imply summation which is always signalled by the explicit use of $\sum$.

## 1.2 Contraction to a Vector

We follow our previous discussion [Bell et al. 2000]. Let $\aleph^{\Im}$ be the complexified algebra of quaternions $(q_\mu \mathbf{i}_\mu, q_\mu \in \Im)$ where $\Im$ is the complex field. We define the $\Im$-linear map,

$$\mathbf{F} : \aleph^{\Im} \to \Im^2, \ \ \mathbf{Q} \to \mathbf{Q}\begin{pmatrix} 1 \\ 0 \end{pmatrix}$$

where **Q** is a complex quaternion, and,

$$\mathbf{i}_r = -i\boldsymbol{\sigma}_r, \tag{1.2.A}$$

Then,

$$\mathbf{F}(\mathbf{QU}) = \mathbf{QF}(\mathbf{U}) \tag{1.2.B}$$

where **U** is a complex quaternion, and the following further properties may be derived,

$$\mathbf{F}(\mathbf{Q}(-\mathbf{i}_3)) = i\mathbf{F}(\mathbf{Q}), \quad \mathbf{F}(\mathbf{Q}(1 + i\mathbf{i}_3)) = 2\mathbf{F}(\mathbf{Q}), \tag{1.2.C}$$

We also define the lift **G**,

$$\mathbf{G} : \Im^2 \to \aleph^{\Im}, \ \ \mathbf{i}_\mu\begin{pmatrix} 1 \\ 0 \end{pmatrix} \to \mathbf{i}_\mu$$

Then,

$$\mathbf{F} \circ \mathbf{G} = id_{\Im^2} \tag{1.2.D}$$

Note that **G** is $\Re$- linear but not $\Im$ - linear where $\Re$ is the field of reals.

The following further properties may be derived for the lift **G**,

$$\mathbf{G}(\mathbf{i}_\mu \mathbf{F}(\mathbf{Q})) = \mathbf{i}_\mu \mathbf{Q}, \ \ \mathbf{G}(i\mathbf{i}_\mu \mathbf{F}(\mathbf{Q})) = \mathbf{i}_\mu \mathbf{Q}(-\mathbf{i}_3) \tag{1.2.E}$$

## 1.3 Contraction to a Scalar

Let,

$$\mathbf{U} = u_0 + \mathbf{i}_1 u_1 + \mathbf{i}_2 u_2 + \mathbf{i}_3 u_3, \tag{1.3.A}$$
$$\mathbf{Q} = q_0 + \mathbf{i}_1 q_1 + \mathbf{i}_2 q_2 + \mathbf{i}_3 q_3$$

where $q_\mu$ and $u_\mu$ may be complex. We call $u_0$, $u_1$, $u_2$ and $u_3$ *the components of* **U**. We define *the dot product of two quaternions*,

$$\mathbf{Q}\cdot\mathbf{U} = u_0 q_0 + u_1 q_1 + u_2 q_2 + u_3 q_3 \tag{1.3.B}$$

As may be easily seen by explicit enumeration,

$$\mathbf{Q}\cdot\mathbf{U} = \left(\mathbf{Q}^{\ddagger}\mathbf{U} + \mathbf{U}^{\ddagger}\mathbf{Q}\right)/2 \tag{1.3.C}$$

where $^{\ddagger}$ signifies quaternion conjugation so that,

$$\mathbf{U}^{\ddagger} = S - \mathbf{V} \tag{1.3.D}$$

where,

$$S = u_0,\quad \mathbf{V} = \mathbf{i}_1 u_1 + \mathbf{i}_2 u_2 + \mathbf{i}_3 u_3, \tag{1.3.E}$$
$$|\,\mathbf{U}\,|^2 = \mathbf{U}^{\ddagger}\mathbf{U},\quad \mathbf{U}^{-1} = \mathbf{U}^{\ddagger}/|\,\mathbf{U}\,|^2$$

We call S *the temporal part* of **U** and **V** *the spatial part*.

We then have for a further contraction operation,

$$\text{Real part of}\left\{[\mathbf{F}(\mathbf{Q})]^{\dagger}\mathbf{i}_r\mathbf{F}(\mathbf{U})\right\} = \mathbf{Q}\cdot(\mathbf{i}_r\mathbf{U}) \tag{1.3.F}$$

from equations (1.2.B) and (1.3.B). Hence from equation (1.3.C),

$$\text{Real part of}\left\{[\mathbf{F}(\mathbf{Q})]^{\dagger}\mathbf{i}_r\mathbf{F}(\mathbf{U})\right\} = \left(\mathbf{Q}^{\ddagger}\mathbf{i}_r\mathbf{U} + \mathbf{U}^{\ddagger}\mathbf{i}_r^{\ddagger}\mathbf{Q}\right)/2 \tag{1.3.G}$$

We now consider a second contraction operation using a Pauli matrix, $\boldsymbol{\sigma}_r$,

$$\text{Real part of}\left\{[\mathbf{F}(\mathbf{Q})]^{\dagger}\boldsymbol{\sigma}_r\mathbf{F}(\mathbf{U})\right\} = \mathbf{Q}\cdot\{\mathbf{i}_r\mathbf{U}(-\mathbf{i}_3)\} \tag{1.3.H}$$

from equations (1.2.A), (1.2.B), (1.2.E) and (1.3.B). Therefore,

$$\text{Real part of}\left\{[\mathbf{F}(\mathbf{Q})]^{\dagger}\boldsymbol{\sigma}_r\mathbf{F}(\mathbf{U})\right\} \tag{1.3.I}$$
$$= \left\{\mathbf{Q}^{\ddagger}\mathbf{i}_r\mathbf{U}(-\mathbf{i}_3) + (-\mathbf{i}_3)^{\ddagger}\mathbf{U}^{\ddagger}\mathbf{i}_r^{\ddagger}\mathbf{Q}\right\}/2$$

from equation (1.3.C).

## 2. DERIVATION OF THE TRIM DIRAC EQUATION

### 2.1 The Dirac Equation

We vary our previous discussion [Bell et al. 2000] of the Dirac equation, which gives the behaviour of a particle in an electromagnetic or gravitational [Bell and Diaz 2004], [Bell 2004] field, by choosing the version given by Dirac [1958] as *the original Dirac equation*,

$$i(\partial\Psi/\partial x_0)=\left\{\sum_r (i\alpha_r(\partial/\partial x_r)+\alpha_r A_r)-A_0+m\beta\right\}\Psi \qquad (2.1.A)$$

where $\Psi$ is the wave function and a column matrix, $x_0$ the temporal co-ordinate, $A$ the potential four-vector,

$$A=(A_0, A_1, A_2, A_3) \qquad (2.1.B)$$

and $m$ the rest mass of the particle. We define the bispinors by,

$$\Psi=(\psi_1,\psi_2)^T \qquad (2.1.C)$$

and the matrices by,

$$\alpha_r=\begin{pmatrix}0 & \sigma_r\\ \sigma_r & 0\end{pmatrix},\ \beta=\begin{pmatrix}1 & 0\\ 0 & -1\end{pmatrix} \qquad (2.1.D)$$

We make the following changes of variable,

$$m^{\sim}=im,\quad A_0^{\sim}=iA_0, \qquad (2.1.E)$$

$$\Phi=\begin{pmatrix}\phi_1\\ \phi_2\end{pmatrix}=\begin{pmatrix}i & -i\\ 1 & 1\end{pmatrix}\begin{pmatrix}\psi_1\\ \psi_2\end{pmatrix}$$

We define *the bijection* **H** between vectors in $\Im^4$ and quaternions,

$$(v_0,v_1,v_2,v_3)\leftrightarrow(v_0+\mathbf{i}_1v_1+\mathbf{i}_2v_2+\mathbf{i}_3v_3)$$

We call the left-hand-side *the vector associated with the quaternion* on the right-hand side. Let,

$$\mathbf{D}^{\sim} = \mathbf{H}\{(i\partial/\partial x_0, \partial/\partial x_1, \partial/\partial x_2, \partial/\partial x_3)\},$$
$$\mathbf{A}^{\sim} = \mathbf{H}\{(A_0^{\sim}, A_1, A_2, A_3)\} \tag{2.1.F}$$

Then equation (2.1.A) becomes

$$(\mathbf{D}^{\sim} - i\mathbf{A}^{\sim})^{\ddagger}\phi_1 = m^{\sim}\phi_2, \quad (\mathbf{D}^{\sim} - i\mathbf{A}^{\sim})\phi_2 = -m^{\sim}\phi_1 \tag{2.1.G}$$

We use the lift **G** of section 1.2,

$$\mathbf{G}\{(\mathbf{D}^{\sim} - i\mathbf{A}^{\sim})^{\ddagger}\phi_1 - m^{\sim}\phi_2\} = 0,$$
$$\mathbf{G}\{(\mathbf{D}^{\sim} - i\mathbf{A}^{\sim})\phi_2 + m^{\sim}\phi_1\} = 0 \tag{2.1.H}$$

From the lift **G** and equations (1.2.E), (2.1.E) and (2.1.F) we obtain for the first of equations (2.1.H),

$$\mathbf{G}(\partial\phi_1/\partial x_0)(-\mathbf{i}_3) + \sum_r(-\mathbf{i}_r\mathbf{G}(\partial\phi_1/\partial x_r)) + A_0\mathbf{G}(\phi_1)$$
$$+\sum_r(\mathbf{i}_r A_r\mathbf{G}(\phi_1)(-\mathbf{i}_3)) - m\mathbf{G}(\phi_2)(-\mathbf{i}_3) = 0 \tag{2.1.I}$$

and a similar equation for the second of equations (2.1.H). We postmultiply both the equations so derived by $(1 + i\mathbf{i}_3)$. We obtain, again using equations (2.1.E) and (2.1.F),

$$(\mathbf{D}^{\sim} - i\mathbf{A}^{\sim})^{\ddagger}\boldsymbol{\phi}_1^{\S} = \boldsymbol{\phi}_2^{\S}\mathbf{M}^{\sim}, \quad (\mathbf{D}^{\sim} - i\mathbf{A}^{\sim})\boldsymbol{\phi}_2^{\S} = -\boldsymbol{\phi}_1^{\S}\mathbf{M}^{\sim\ddagger} \tag{2.1.J}$$

where,

$$\mathbf{M}^{\sim} = \mathbf{M}^{\sim\ddagger} = m^{\sim} \tag{2.1.K}$$

and we choose to write $\mathbf{M}^{\sim}$ in deference to properties we will discuss in section 3.3, and,

$$\phi_1 = \mathbf{G}(\phi_1), \qquad \phi_2 = \mathbf{G}(\phi_2),$$
$$\phi_1^{\S} = \phi_1(1 + i\mathbf{i}_3), \quad \phi_2^{\S} = \phi_2(1 + i\mathbf{i}_3) \tag{2.1.L}$$

Equations (2.1.L) define a new form of the bispinors. We may map equations (2.1.J) with **F**,

$$\mathbf{F}\left\{\left(\mathbf{D}^{\sim} - i\mathbf{A}^{\sim}\right)^{\ddagger}\phi_1^{\S} - \mathbf{M}^{\sim}\phi_2^{\S}\right\} = 0,$$
$$\mathbf{F}\left\{\left(\mathbf{D}^{\sim} - i\mathbf{A}^{\sim}\right)\phi_2^{\S} + \mathbf{M}^{\sim\ddagger}\phi_1^{\S}\right\} = 0 \tag{2.1.M}$$

and from the definition of map **F** and equations (1.2.B) and (1.2.C) or (1.2.D) we obtain equations (2.1.G). Thus, while $\mathbf{M}^{\sim}$ remains a scalar equal to $m^{\sim}$, the Dirac equation, (2.1.J), is simply another form of the original equation.

The Dirac equation in the bispinors becomes the same equation in the spinor reflector, $\underline{\mathbf{\Phi}}^{\S}\left(\phi_1^{\S}, \phi_2^{\S}\right)$,

$$\left(\underline{\mathbf{D}}^{\sim}\left(\mathbf{D}^{\sim}, \mathbf{D}^{\sim\ddagger}\right) - i\underline{\mathbf{A}}^{\sim}\left(\mathbf{A}^{\sim}, \mathbf{A}^{\sim\ddagger}\right)\right)\underline{\mathbf{\Phi}}^{\S}\left(\phi_1^{\S}, \phi_2^{\S}\right) =$$
$$\underline{\mathbf{\Phi}}^{\S}\left(\phi_1^{\S}, \phi_2^{\S}\right)\underline{\mathbf{M}}^{\sim}\left(\mathbf{M}^{\sim}, -\mathbf{M}^{\sim\ddagger}\right)$$

or,

$$\left(\underline{\mathbf{D}}^{\sim} - i\underline{\mathbf{A}}^{\sim}\right)\underline{\mathbf{\Phi}}^{\S} = \underline{\mathbf{\Phi}}^{\S}\,\underline{\mathbf{M}}^{\sim} \tag{2.1.N}$$

## 2.2 The Trim Dirac Equation

We may write the Dirac equation as,

$$\left\{\left(\underline{\mathbf{D}}^{\sim} - i\underline{\mathbf{A}}^{\sim} - \mathbf{M}^{\sim}|\right)\boldsymbol{\Phi}\right\}(1 + i\mathbf{i}_3) = 0 \tag{2.2.A}$$

where,

$$\mathbf{M}^{\sim}| = \mathbf{M}^{\sim}|\left(-\mathbf{M}^{\sim\ddagger}, \mathbf{M}^{\sim}\right), \quad \boldsymbol{\Phi} = \begin{pmatrix} \phi_1 \\ \phi_2 \end{pmatrix} \tag{2.2.B}$$

from equations (2.1.K), (2.1.L) and (2.1.N). The right-hand side will be zero if one of two conditions are met,

(1) $$\left(\underline{\mathbf{D}}^{\sim} - i\underline{\mathbf{A}}^{\sim} - \mathbf{M}^{\sim}|\right)\boldsymbol{\Phi} = 0, \tag{2.2.C}$$

(2) $$\left(\underline{\mathbf{D}}^{\sim} - i\underline{\mathbf{A}}^{\sim} - \mathbf{M}^{\sim}|\right)\boldsymbol{\Phi} = X\left(1 - i\mathbf{i}_3\right)$$

where $X$ is a column matrix with quaternion elements, since, if $\mathbf{Y}\left(1 + i\mathbf{i}_3\right) = 0$, where $\mathbf{Y}$ is a quaternion, $\mathbf{Y}$ must be either $0$ or $\mathbf{Y}'\left(1 - i\mathbf{i}_3\right)$, where $\mathbf{Y}'$ may be any quaternion. Since,

$$\mathbf{F}\left\{X\left(1 - i\mathbf{i}_3\right)\right\} = 0 \tag{2.2.D}$$

from the definition of map $\mathbf{F}$ and equations (1.2.B) and (1.2.C), we obtain the same copy of the original Dirac equation whatever the value of $X$ and we may set it to any convenient value. We choose $X = 0$ so that the second equation is included in the first. This abolishes the special position that $\mathbf{i}_3$ has held hitherto. The result is

$$\left(\underline{\mathbf{D}}^{\sim} - i\underline{\mathbf{A}}^{\sim}\right)\underline{\mathbf{\Phi}} = \underline{\mathbf{\Phi}}\,\underline{\mathbf{M}}^{\sim} \tag{2.2.E}$$

Let the temporal part of a reflector, consisting of the temporal parts of the constituent quaternions, be indicated by the subscript $(t')$, while the corresponding spatial part is indicated by $(s)$. Resolving the equation into real and imaginary parts,

$$\begin{gathered} \left(\underline{\mathbf{D}}^{\sim}{}_{(s)} - i\underline{\mathbf{A}}^{\sim}{}_{(t')}\right)\underline{\mathbf{\Phi}} = \underline{\mathbf{\Phi}}\,\underline{\mathbf{M}}^{\sim}{}_{(s)}, \\ \left(\underline{\mathbf{D}}^{\sim}{}_{(t')} - i\underline{\mathbf{A}}^{\sim}{}_{(s)}\right)\underline{\mathbf{\Phi}} = \underline{\mathbf{\Phi}}\,\underline{\mathbf{M}}^{\sim}{}_{(t')} \end{gathered} \tag{2.2.F}$$

from the definitions of the variables in section 2.1, where we have allowed $\underline{\mathbf{M}}^{\sim}$ the spatial part discussed in section 3.3. Setting $\underline{\mathbf{X}}^{\sim}_{(t)} = i\underline{\mathbf{X}}^{\sim}_{(t')}$, the equations become,

$$\begin{aligned} \left(\underline{\mathbf{D}}^{\sim}{}_{(s)} - \underline{\mathbf{A}}^{\sim}{}_{(t)}\right)\underline{\mathbf{\Phi}} &= \underline{\mathbf{\Phi}}\,\underline{\mathbf{M}}^{\sim}{}_{(s)}, \\ \left(\underline{\mathbf{D}}^{\sim}{}_{(t)} + \underline{\mathbf{A}}^{\sim}{}_{(s)}\right)\underline{\mathbf{\Phi}} &= \underline{\mathbf{\Phi}}\,\underline{\mathbf{M}}^{\sim}{}_{(t)} \end{aligned} \tag{2.2.G}$$

where all the variables are real. These two equations are yet another form of the original Dirac equation, in that using the map **G** we may map the original into this form and, using the map **F** we may map these equations into the original Dirac equation. If we allow every individual part to transform relativistically we could use this form in what we say about the Dirac equation in the rest of the paper. However, we may also produce a single equation, which is the generalisation of both and overtly relativistic. Let,

$$\begin{aligned} &\mathbf{D} = \mathbf{H}\{(\partial/\partial x_0,\ \partial/\partial x_1,\ \partial/\partial x_2,\ \partial/\partial x_3)\},\ \ \underline{\mathbf{D}} = \underline{\mathbf{D}}(\mathbf{D}, \mathbf{D}^{\ddagger}) \\ &\mathbf{A} = \mathbf{H}\{(\mathbf{A}_0, \mathbf{A}_1, \mathbf{A}_2, \mathbf{A}_3)\},\ \ \underline{\mathbf{A}} = \underline{\mathbf{A}}(\mathbf{A}, \mathbf{A}^{\ddagger}), \\ &|\mathbf{A}_0| = A_0,\ \ |(\mathbf{A}_1, \mathbf{A}_2, \mathbf{A}_3)| = -|(A_1, A_2, A_3)|, \\ &\mathbf{M} = -i\mathbf{M}^{\sim},\ \ \underline{\mathbf{M}} = (\mathbf{M}, -\mathbf{M}^{\ddagger}) \end{aligned} \tag{2.2.H}$$

then we may rewrite equation (2.2.G) as,

$$\begin{aligned} \left(\underline{\mathbf{D}}_{(s)} - \underline{\mathbf{A}}_{(t)}\right)\underline{\mathbf{\Phi}} &= \underline{\mathbf{\Phi}}\,\underline{\mathbf{M}}_{(s)}, \\ \left(\underline{\mathbf{D}}_{(t)} - \underline{\mathbf{A}}_{(s)}\right)\underline{\mathbf{\Phi}} &= \underline{\mathbf{\Phi}}\,\underline{\mathbf{M}}_{(t)} \end{aligned} \tag{2.2.I}$$

Setting $\underline{\mathbf{M}}_{(s)} = 0$, the first equation shows that $\underline{\mathbf{D}}_{(s)}\underline{\mathbf{\Phi}} = \underline{\mathbf{A}}_{(t)}\underline{\mathbf{\Phi}}$. This is surprising. We will allow the quaternion in $\underline{\mathbf{A}}_{(t)}$ to have a zero temporal part but spatial parts depending on the $\mathbf{i}_r$, while conversely, the quaternion in $\underline{\mathbf{A}}_{(s)}$ is a scalar. With this in mind, we have made the components of **A** themselves quaternions. We demonstrate the effect in the next section. Adding the two equations,

$$\left(\underline{\mathbf{D}} - \underline{\mathbf{A}}\right)\underline{\boldsymbol{\Phi}} = \underline{\boldsymbol{\Phi}}\,\underline{\mathbf{M}} \tag{2.2.J}$$

This is the trim Dirac equation. However, if we want to be sure that a solution of the trim Dirac equation is a solution of the original Dirac equation, we must insist that it also obeys equations (2.2.G).

We remark that the difficulty with **A** which emerges in the derivation of the field version [Greiner and Reinhardt 1996b] leading to the appearance of "ghosts" [Kaku 1993] may be connected to some of the properties of the Dirac equation found in this section.

## 2.3 A Simple Solution of the Trim Dirac Equation

We will find the solution we used throughout our previous work [Bell et al. 2004a&b], [Bell and Diaz 2004], [Bell 2004] in which the potential is constant. Let,

$$\begin{aligned} \phi_2 &= \mathbf{X}\sin\left(\nu x_0 + \mu_r x_r\right) + \mathbf{Y}\cos\left(\nu x_0 + \mu_r x_r\right) \\ \phi_1 &= \mathbf{B}\sin\left(\nu x_0 + \mu_r x_r\right) + \mathbf{C}\cos\left(\nu x_0 + \mu_r x_r\right) \end{aligned} \tag{2.3.A}$$

Writing equation (2.2.J) in terms of the bispinors,

$$\left(\mathbf{D} - \mathbf{A}\right)\phi_2 = -\phi_1\mathbf{M}^{\ddagger}, \quad \left(\mathbf{D} - \mathbf{A}\right)^{\ddagger}\phi_1 = \phi_2\mathbf{M} \tag{2.3.B}$$

and applying the first of equations (2.3.A) to the first of (2.3.B) and the same for the second,

$$\phi_1 = -\{\nu \mathbf{X} + \mathbf{i}_r \mu_r \mathbf{X} - \mathbf{A}\mathbf{Y}\}\cos(\nu x_0 + \mu_r x_r)\mathbf{M}/m^2 + \{\nu \mathbf{Y} + \mathbf{i}_r \mu_r \mathbf{Y} + \mathbf{A}\mathbf{X}\}\sin(\nu x_0 + \mu_r x_r)\mathbf{M}/m^2$$
$$\phi_2 = \{\nu \mathbf{B} - \mathbf{i}_r \mu_r \mathbf{B} - \mathbf{A}^\ddagger \mathbf{C}\}\cos(\nu x_0 + \mu_r x_r)\mathbf{M}^\ddagger/m^2 - \{\nu \mathbf{C} - \mathbf{i}_r \mu_r \mathbf{C} + \mathbf{A}^\ddagger \mathbf{B}\}\sin(\nu x_0 + \mu_r x_r)\mathbf{M}^\ddagger/m^2 \quad (2.3.C)$$

The expressions for $\phi_1$ and $\phi_2$ in equations (2.3.A) and (2.3.C) must be equal for all values of $x_\mu$ if both are to represent a solution, which means that,

$$\mathbf{C} = -(\nu \mathbf{X} + \mathbf{i}_r \mu_r \mathbf{X} - \mathbf{A}\mathbf{Y})\mathbf{M}/m^2$$
$$\mathbf{B} = (\nu \mathbf{Y} + \mathbf{i}_r \mu_r \mathbf{Y} + \mathbf{A}\mathbf{X})\mathbf{M}/m^2$$
$$\mathbf{Y} = (\nu \mathbf{B} - \mathbf{i}_r \mu_r \mathbf{B} - \mathbf{A}^\ddagger \mathbf{C})\mathbf{M}^\ddagger/m^2$$
$$\mathbf{X} = -(\nu \mathbf{C} - \mathbf{i}_r \mu_r \mathbf{C} + \mathbf{A}^\ddagger \mathbf{B})\mathbf{M}^\ddagger/m^2 \quad (2.3.D)$$

Substituting for **C** and **B** from the first two equations into the last two,

$$(m^2 - \nu^2 - \mu_r^2 + \mathbf{A}^\ddagger \mathbf{A})\mathbf{Y} = \{(\nu - \mathbf{i}_r \mu_r)\mathbf{A} + \mathbf{A}^\ddagger(\nu + \mathbf{i}_r \mu_r)\}\mathbf{X}$$
$$(m^2 - \nu^2 - \mu_r^2 + \mathbf{A}^\ddagger \mathbf{A})\mathbf{X} = -\{(\nu - \mathbf{i}_r \mu_r)\mathbf{A} + \mathbf{A}^\ddagger(\nu + \mathbf{i}_r \mu_r)\}\mathbf{Y} \quad (2.3.E)$$

We allow for this relation by setting,

$$\mathbf{Y} = -\mathbf{i}_r \mathbf{X} \quad (2.3.F)$$

Then, provided the subscript *r* denotes a single term,

$$m^2 - \nu^2 - \mu_r^2 + \mathbf{A}^\ddagger \mathbf{A} - 2\nu \mathbf{A}_0 \mathbf{i}_r - 2\mu_r \mathbf{A}_r \mathbf{i}_r = 0 \quad (2.3.G)$$

or,

$$m^2 + (\mathbf{i}_r \nu - \mathbf{A}_0)^2 + (\mathbf{i}_r \mu_r - \mathbf{A}_r)^2 = 0 \quad (2.3.H)$$

We use the definitions in equations (2.2.H), and set,

$$\mathbf{A}_0 = \mathbf{i}_r A_0, \quad \mathbf{A}_r = -\mathbf{i}_r A_r \tag{2.3.I}$$

at which point, on making the temporal or spatial variables imaginary, we retrieve the same solution as we did previously for the Dirac equation [Bell et al. 2004a], that is, with the conventions used here,

$$m^2 = (\nu + A_0)^2 + (\mu_r - A_r)^2 \tag{2.3.J}$$

Using equations (2.3.F) and (2.3.D), we may modify equations (2.3.A) to read,

$$\begin{aligned} \phi_2 &= \mathbf{Y} \exp\{\mathbf{i}_r (\nu x_0 + \mu_r x_r)\} \\ \phi_1 &= \mathbf{Y}\{-\mathbf{i}_r \nu + \mu_r + \mathbf{A}\} \mathbf{M}/m^2 \exp\{\mathbf{i}_r (\nu x_0 + \mu_r x_r)\} \end{aligned} \tag{2.3.K}$$

We may discover the more general representation for multiple values of $r$ by spatial rotation, which we discuss in section 3.

## 2.4 Series Solution of the Trim Dirac Equation

We derive the propagator for the trim Dirac equation following Greiner and Reinhardt [1996a]. We may choose any reference frame and transform equation (2.2.J) using equations (2.2.B) and (2.2.H) to find,

$$(\underline{\mathbf{D}} - \underline{\mathbf{A}})\boldsymbol{\Phi} = \mathbf{M} | (-\mathbf{M}^{\ddagger}, \mathbf{M})\boldsymbol{\Phi} \tag{2.4.A}$$

We will change the name of the propagator, $S_F$, to $\mathbf{B}_\mathbf{D}$ in honour of Bohr and Dirac since ours differs [Greiner & Reinhardt 1996a], [Bell et al. 2004a&b]. It is given by,

$$(\underline{\mathbf{D}}[x'] - \underline{\mathbf{A}}[x'] - \mathbf{M}|)\, \mathbf{B}_\mathbf{D}[x' - x; \mathbf{A}] = \delta^4[x' - x] \tag{2.4.B}$$

where the brackets, [ ], indicate a dependence on Cartesian co-ordinates.

The free propagator is given by,

$$\left(\underline{\mathbf{D}}[x'] - \mathbf{M}\mathbb{I}\right)\mathbf{B_D}[x'-x] = \delta^4[x'-x] \qquad (2.4.C)$$

We transform to momentum space,

$$\mathbf{B_D}[x'-x] = \int \frac{\mathrm{d}^4 p}{(2\pi)^4} \exp\{-ip \cdot (x'-x)\} \mathbf{B_D}[p] \qquad (2.4.D)$$

where $p$ is the energy-momentum four-vector and $\mathbf{B_D}[p]$ is the free propagator in momentum space. Inserting equation (2.4.D) into equation (2.4.C), expanding the δ function elements on the right hand side by,

$$\delta[x'-x] = \int \frac{\mathrm{d}^4 p}{(2\pi)^4} \exp\{-ip \cdot (x'-x)\} \qquad (2.4.E)$$

and performing the differentiation,

$$\left(\underline{\mathbf{P}} - \mathbf{M}\mathbb{I}\right)\mathbf{B_D}[p] = 1 \qquad (2.4.F)$$

where,

$$\underline{\mathbf{P}} = \underline{\mathbf{P}}\left(\mathbf{P}, \mathbf{P}^{\ddagger}\right), \quad \mathbf{P} = \sum_{\mu} p_{\mu} \mathbf{i}_{\mu} \qquad (2.4.G)$$

Equation (2.4.F) is satisfied by,

$$\mathbf{B_D}[p] = \frac{\underline{\mathbf{P}} + \mathbf{M}\mathbb{I}}{p^2 - m^2} \qquad (2.4.H)$$

We may write,

$$\left(\underline{\mathbf{D}}[x'] - \mathbf{M}\mathbb{I}\right)\underline{\boldsymbol{\Phi}}[x'] = \boldsymbol{\chi}[x'] \qquad (2.4.I)$$

where,

$$\boldsymbol{\chi}[x'] = \delta^4[x'-x] + \underline{\mathbf{A}}[x']\mathbf{B_D}[x'-x; \mathbf{A}] \qquad (2.4.J)$$

and we interpret $\boldsymbol{\chi}$ as a source term. The solution is given by,

$$\Phi[x'] = \int d^4 y \mathbf{B_D}[x'-y]\chi[y] \qquad (2.4.K)$$

Replacing $\Phi[x']$ by $\mathbf{B_D}[x'-x;\mathbf{A}]$ and substituting for $\chi[y]$ from equation (2.4.J) leads to,

$$\mathbf{B_D}[x'-x;\mathbf{A}] = \mathbf{B_D}[x'-x] + \int d^4 y \mathbf{B_D}[x'-y]\underline{\mathbf{A}}[y]\mathbf{B_D}[y-x;\mathbf{A}] \qquad (2.4.L)$$

which is the Lippmann-Schwinger equation [Greiner and Reinhardt 1996a]. The Dirac equation with a source term is not explicitly covariant for four-vector behaviour of the mass term but we may find the wave function from the propagator, which does satisfy the Dirac equation with joint four-vector behaviour of the wave function and mass term.

# 3. SPATIAL AND TEMPORAL ROTATIONS

## 3.1 Lorentz Transformations

Let $Q$ be the four-vector,

$$Q = (q_0, q_1, q_2, q_3) \qquad (3.1.A)$$

with $q_\mu$ real, in Minkowski spacetime with signature $-+++$. Any Lorentz transformation of $Q$ may be represented by a Lorentz transformation confined to the $(q_0, q_1)$ plane combined with spatial rotations. Without loss of generality we consider only the former. Such a Lorentz transformation from the undashed to the dashed frame, travelling at a velocity $-v$ along the $q_1$-axis, may be described by,

$$\begin{pmatrix} q_0' \\ q_1' \end{pmatrix} = \begin{pmatrix} \cosh\theta & \sinh\theta \\ \sinh\theta & \cosh\theta \end{pmatrix} \begin{pmatrix} q_0 \\ q_1 \end{pmatrix}, \qquad (3.1.B)$$

$$\sinh\theta = \frac{v}{\sqrt{1-v^2}}, \quad \cosh\theta = \frac{1}{\sqrt{1-v^2}}$$

where $\theta$ is real [Elliott and Dawber 1979a&b]. We shall call this *the Minkowski transformation*. Setting,

$$\cos i\theta = \cosh\theta, \quad \sin i\theta = i\sinh\theta \qquad (3.1.C)$$

equation (3.1.B) becomes,

$$\begin{pmatrix} iq_0' \\ q_1' \end{pmatrix} = \begin{pmatrix} \cos i\theta & -\sin i\theta \\ \sin i\theta & \cos i\theta \end{pmatrix} \begin{pmatrix} iq_0 \\ q_1 \end{pmatrix} \qquad (3.1.D)$$

which is the basis for our previous treatment of Lorentz transformations [Bell et al. 2000]. We may instead represent equation (3.1.B) using a different parameterisation,

$$\sec\phi = \cosh\theta, \quad \tan\phi = \sinh\theta \qquad (3.1.E)$$

in which the Minkowski transformation becomes,

$$\begin{pmatrix} q_0' \\ q_1' \end{pmatrix} = \begin{pmatrix} \sec\phi & \tan\phi \\ \tan\phi & \sec\phi \end{pmatrix} \begin{pmatrix} q_0 \\ q_1 \end{pmatrix} \qquad (3.1.F)$$

$$\sin\phi = v, \quad \cos\phi = \sqrt{1-v^2}$$

where $\phi$ is real. As for equation (3.1.B), the lines $\phi = \pm\pi/4$ and $\phi = \pm 3\pi/4$ represent the light cone, where the value of elements of four-vectors in the dashed frame approach infinity [Elliott and Dawber 1979b]. The regions $|\phi| < \pi/4$ are physically accessible, while the regions $|\phi| > \pi/4$ describe tachyons. The equation leads to,

$$\begin{pmatrix} q_0 \\ q_1' \end{pmatrix} = \begin{pmatrix} \cos\phi & -\sin\phi \\ \sin\phi & \cos\phi \end{pmatrix} \begin{pmatrix} q_0' \\ q_1 \end{pmatrix} \tag{3.1.G}$$

where we have made the exchange,

$$iq_0' \to q_0, \;\; iq_0 \to q_0', \;\; i\theta \to \phi \tag{3.1.H}$$

in going from equation (3.1.D) to (3.1.G). The stationary and the moving frames are now represented by the vectors $(q_0', q_1)^T$ and $(q_0, q_1')^T$ rather than $(q_0, q_1)^T$ and $(q_0', q_1')^T$. We will call this *the Euclidean transformation*. It applies to a Euclidean spacetime, where the signature is $++++$. By contrast with the Minkowski transformation in equation (3.1.F), the light cone has opened into a semi-hypersphere, represented here by the line $\phi = \pm\pi/2$, and there is no discontinuity. The elements of the four-vector in the moving frame remain finite while the direction of travel along the observer's temporal axis reverses. The tachyonic portions of the graph have been excised, and radiation travels infinitely fast. Since the Galilean transformations [Einstein 1958] assume the temporal co-ordinates in the two frames are identical, it is no easier to choose the route Einstein chose to Special Relativity than to regard equation (3.1.G) as defining Lorentz transformations. However, as is well-known, the Maxwell equations embody the usual transformation, equation (3.1.B) or (3.1.F), and so it is Maxwell not Einstein who made the decision. We now tend to regard the temporal co-ordinate as "owned" by the observer and the spatial co-ordinate as "shared" between observers and it is also possible to arrive at a similar formulation using the map,

$$iq_1' \to q_1, \;\; iq_1 \to q_1', \;\; i\theta \to \phi \tag{3.1.I}$$

in which a spatial co-ordinate is exchanged.

Transformation of a Lorentz invariant equation from Minkowski spacetime to Euclidean spacetime, as defined here, requires the knowledge of the equation in two different frames. We may however reduce this to one frame by considering the limit $\theta \to 0,$ implying $\phi \to 0,$ which leaves the matrices in equations (3.1.D) and (3.1.G) equal. We may then freely exchange $q_0$ and $q_0'$. This means we may consider the behaviour of the Dirac equation under a Lorentz transformation as given by either equation (3.1.B) and (3.1.D) or (3.1.G). The same applies to the radiation equation, which describes the electromagnetic or gravitational [Bell and Diaz 2004a&b] field generated by a charged particle. We may also, with more economy represent both forms of transformation as,

$$\begin{pmatrix} \tilde{q}_0' \\ \tilde{q}_1' \end{pmatrix} = \begin{pmatrix} \cos\tilde{\phi} & -\sin\tilde{\phi} \\ \sin\tilde{\phi} & \cos\tilde{\phi} \end{pmatrix} \begin{pmatrix} \tilde{q}_0 \\ \tilde{q}_1 \end{pmatrix} \quad (3.1.J)$$

where we allow all the variables to be real or all real except for an imaginary angle $\tilde{\phi}$ and imaginary temporal co-ordinates.

Suppose that an equation between four-vectors holds whether the first element of the four-vector of interest, here represented by $q_0$, is given a real or an imaginary value. Suppose further that the equation can be shown to be invariant under the operation given in equation (3.1.J) by a proof that does not depend on whether $\tilde{\phi}$ has a real or imaginary value. Then the proof of invariance for the Euclidean transformation is sufficient to ensure invariance for the Minkowski and vice versa. This is true for both the Dirac and the radiation equation as we will show. Since we may make the transition between the two types of invariance, Minkowski and Euclidean, radiation travelling at the speed of light and a ray which travels infinitely fast may be represented within the one spacetime by two copies of the Dirac and radiation equations, transforming in different ways. We will call the former radiation *terrestrial*, and represent it on the light cone at an angle of $\pi/4$ to the axes. We will call the latter, *celestial*, and represent it on *the celestial plane* along the spatial axes. The difference is a little like that between covariant and contravariant vectors, when curvature is used to describe gravity [Einstein 1967]. It was found that a distinguishing notation was needed for these, and in practice we shall need one here, if this prediction of the equations proves to have physical explanatory value. We shall discuss such a notation at the end of the next section. We remark that most of the theoretical predictions of QED theory have been found to have physical counterparts in the past.

### 3.2 Spatial and Temporal Rotations of Reflectors and Rotators

We introduce practical methods of representing the fundamental rotations just discussed. Associating four-vectors with a quaternion using the map **H**, we may describe their spatial rotation and Lorentz transformation in two ways. In the first we transform the components of the quaternions, identically the elements of the Euclidean four-vectors, using a suitable spatial rotation or boost matrix and leave the quaternion matrices, $\mathbf{i}_\mu$, constant. This is the approach we used in the last section. In the second we leave the components of the quaternion constant and transform the quaternion matrices, $\mathbf{i}_\mu$, using quaternion multiplication. Dirac [Dirac 1958] chose the former course, while from now on we will choose the latter.

We have studied the geometry of quaternion and hence reflector and rotator multiplications before [Bell and Mason 1990], [Bell et al. 2000]. We recapitulate and augment. Let **R** be a quaternion with complex components and unit modulus with,

$$\mathbf{R} = r_0 + \mathbf{i}_1 r_1 + \mathbf{i}_2 r_2 + \mathbf{i}_3 r_3 \qquad (3.2.A)$$

and,

$$S_R = \mathbf{H}^{-1}(r_0) = (r_0, 0, 0, 0), \qquad (3.2.B)$$
$$V_R = \mathbf{H}^{-1}(\mathbf{i}_1 r_1 + \mathbf{i}_2 r_2 + \mathbf{i}_3 r_3) = (0, r_1, r_2, r_3)$$

We call the plane containing $S_R$ and $V_R$ *the temporal plane* and the plane everywhere at right angles to it *the spatial plane*. We define,

$$\tan(\phi/2) = \sqrt{(r_1^2 + r_2^2 + r_3^2)}\Big/ r_0 \qquad (3.2.C)$$

Let **Q** be a quaternion with complex components. Table I gives the angular behaviour of $\mathbf{H}^{-1}(\mathbf{Q})$ under quaternion multiplication by **R**. The first column defines the type of multiplication and the second two columns define the angle, $\xi$, through which **Q** turns to become $\mathbf{Q}'$. The first column gives the angle, $\xi_s$, in the spatial plane and the second the angle, $\xi_t$, in the temporal plane.

**Table I.** The Rotations Effected by Quaternion Multiplication

| Expression, $\mathbf{Q}' =$ | Spatial plane $\xi_s$ | Temporal plane $\xi_t$ |
|---|---|---|
| $\mathbf{RQ}$ | $+\xi/2$ | $+\xi/2$ |
| $\mathbf{QR}$ | $-\xi/2$ | $+\xi/2$ |
| $\mathbf{R}^{\ddagger}\mathbf{Q}$ | $-\xi/2$ | $-\xi/2$ |
| $\mathbf{QR}^{\ddagger}$ | $+\xi/2$ | $-\xi/2$ |
| $\mathbf{RQR}$ | $0$ | $+\xi$ |
| $\mathbf{RQR}^{\ddagger}$ | $+\xi$ | $0$ |
| $\mathbf{R}^{\ddagger}\mathbf{QR}$ | $-\xi$ | $0$ |
| $\mathbf{R}^{\ddagger}\mathbf{QR}^{\ddagger}$ | $0$ | $-\xi$ |

We call a rotation in the spatial plane *a spatial rotation* and a rotation in the temporal plane *a temporal rotation.* The latter represents a boost or Lorentz transformation.

So, for example, the Euclidean transformation described in equation (3.1.G) would be represented by,

$$\mathbf{R} = \cos\frac{\phi}{2} + \mathbf{i}_1 \sin\frac{\phi}{2} \qquad (3.2.D)$$

and,

$$\mathbf{Q}' = \mathbf{RQR} \qquad (3.2.E)$$

with $\phi$ and the components of $\mathbf{Q}$ and $\mathbf{Q}'$ real, while the Minkowski transformation in equation (3.1.D) would be represented by the same but with,

$$\mathbf{R} = \cos\frac{i\theta}{2} + \mathbf{i}_1 \sin\frac{i\theta}{2} \tag{3.2.F}$$

and the spatial part of $\mathbf{Q}$ and $\mathbf{Q}'$ real but the temporal part imaginary. Here $\phi$ and $\theta$ have the same relationship as in the last section.

We develop the properties of the reflector matrices under rotation. We consider the quaternion equation,

$$\mathbf{QP}^{\ddagger} = \mathbf{PW}^{\ddagger} \tag{3.2.G}$$

We may provide the four-vectors $\mathbf{H}^{-1}(\mathbf{Q})$, $\mathbf{H}^{-1}(\mathbf{P})$ and $\mathbf{H}^{-1}(\mathbf{W})$ with a spatial rotation by pre- and post-multiplication of the quaternions by a suitable $\mathbf{R}$ and $\mathbf{R}^{\ddagger}$. This transformation is a similarity transformation that leaves equation (3.2.G) invariant. However, if we wish to give the variables a temporal rotation, we must pre- and post-multiply $\mathbf{Q}$ and $\mathbf{P}$ by $\mathbf{R}$ while $\mathbf{P}^{\ddagger}$ and $\mathbf{W}^{\ddagger}$ must be pre- and post-multiplied by $\mathbf{R}^{\ddagger}$ to preserve the relation of quaternion conjugation. Neither of these is a similarity transformation although equation (3.2.G) remains invariant. However, if we take the quaternion conjugate of equation (3.2.G) and re-arrange we obtain a second form of the equation,

$$\mathbf{Q}^{\ddagger}\mathbf{P} = \mathbf{P}^{\ddagger}\mathbf{W} \tag{3.2.H}$$

We may write equations (3.2.G) and (3.2.H) as,

$$\underline{\mathbf{Q}}\left(\mathbf{Q},\mathbf{Q}^{\ddagger}\right)\underline{\mathbf{P}}\left(\mathbf{P},\mathbf{P}^{\ddagger}\right) = \underline{\mathbf{P}}\left(\mathbf{P},\mathbf{P}^{\ddagger}\right)\underline{\mathbf{W}}\left(\mathbf{W},\mathbf{W}^{\ddagger}\right) \tag{3.2.I}$$

Equation (3.2.I) implies no more than equation (3.2.G) but we may now provide a temporal rotation which leaves equation (3.2.I) invariant and is a similarity transformation, as follows. Let,

$$\mathbf{R}_S| = \mathbf{R}_S|(\mathbf{R},\mathbf{R}), \quad \mathbf{R}_T| = \mathbf{R}_T|(\mathbf{R},\mathbf{R}^{\ddagger}) \qquad (3.2.J)$$

Then we may effect a rotation of $\mathbf{H}^{-1}(\mathbf{Q})$, $\mathbf{H}^{-1}(\mathbf{P})$ and $\mathbf{H}^{-1}(\mathbf{W})$ by,

$$\underline{\mathbf{Q}'} = \mathbf{R}|^{a}\underline{\mathbf{Q}}\mathbf{R}|^{\ddagger a}, \quad \underline{\mathbf{P}'} = \mathbf{R}|^{a}\underline{\mathbf{Q}}\mathbf{R}|^{\ddagger b}, \quad \underline{\mathbf{W}'} = \mathbf{R}|^{b}\underline{\mathbf{Q}}\mathbf{R}|^{\ddagger b} \qquad (3.2.K)$$

where $a = b = 1$ provides four-vector behaviour for all the associated four-vectors with $\mathbf{R}| = \mathbf{R}_S|$ for a spatial rotation and $\mathbf{R}| = \mathbf{R}_T|$ for a temporal rotation. Although other values may also be interpreted using the first and fourth entries in table I as a combination of rotations through angles of $\pm a\xi/2$ or $\pm b\xi/2$ [Bell and Mason 1990], we shall confine our discussion to four-vectors for the rest of this section.

We may broaden our remarks and include,

$$\underline{\mathbf{Q}} = \underline{\mathbf{Q}}(\mathbf{Q},\mathbf{U}) \qquad (3.2.L)$$

where $\mathbf{Q}$ and $\mathbf{U}$ are distinct, providing that $\mathbf{U}$ rotates in the same sense as $\mathbf{Q}^{\ddagger}$ under a temporal rotation. Rotations are described by,

$$\underline{\mathbf{Q}'}(\mathbf{Q}',\mathbf{U}') = \mathbf{R}|\,\underline{\mathbf{Q}}(\mathbf{Q},\mathbf{U})\mathbf{R}|^{\ddagger} \qquad (3.2.M)$$

where $\mathbf{R}| = \mathbf{R}_S|$ for a spatial rotation and $\mathbf{R}| = \mathbf{R}_T|$ for a temporal rotation. Further, all the reflectors in equation (3.2.I) may be distinct. Thus equations consisting of reflector multiplications are invariant under any rotation in $\Im^4$ up to a similarity transformation.

Suppose we have an equation in which the product of an even number of reflector matrices, for example, $\underline{\mathbf{P}}(\mathbf{P},\mathbf{S})$, is set equal to a rotator matrix, $\mathbf{Q}|(\mathbf{Q},\mathbf{U})$. For a rotation of the four-vectors associated with the quaternions in the $\underline{\mathbf{P}}(\mathbf{P},\mathbf{S})$, we have for the rotator,

$$\mathbf{Q}|' = \mathbf{R}|\,\mathbf{Q}|\,\mathbf{R}|^{\ddagger} \qquad (3.2.N)$$

which provides a spatial rotation of $\mathbf{Q}$ and $\mathbf{U}$ whether $\mathbf{R}|$ is $\mathbf{R}_S|$ or $\mathbf{R}_T|$. Thus the temporal parts of the quaternions associated with the rotator are constant under any type of rotation.

We have now shown that all spatial rotations and Lorentz transformations of any Euclidean four-vectors associated with $\underline{\mathbf{Q}}$ or $\mathbf{Q}|$ can be effected by suitably chosen $\mathbf{R}_S|$ and $\mathbf{R}_T|$. We may go further. We may use the Dirac equation, (2.2.J), to represent both the terrestrial and the celestial case. Both **M** and **A** have only a temporal part in the rest frame, from equations (2.2.H), where we will suppose the terrestrial and celestial coincide. Provided we choose the local frame in which $\phi_1 = \phi_2$, which we discuss in section 4 and the appendix, and have the liberty to set the phase accordingly, $\phi_1$ and $\phi_2$ conform. Temporal rotations are represented by a complex angle,

$$\gamma = \phi + i\theta \qquad (3.2.O)$$

Subsequently, after a series of Lorentz transformations, performed separately for the Euclidean and Minkowski cases and added, and in the general case, the terrestrial quantities will be represented by the modulus of imaginary variables and the celestial by real variables, provided $\phi_1$ behaves like a four-vector, from equations (3.2.D) through (3.2.F). We leave open whether it were then possible to rotate the real and imaginary plane by differing angles, whether in the spatial or the temporal planes, and the significance of the half-angle rotation permitted $\phi_1$ below.

### 3.3 Invariances of the Trim Dirac Equation

We discuss the behaviour of the trim Dirac equation under spatial and temporal rotation in Euclidean spacetime. Using equation (3.2.M) for the four-vectors,

$$\underline{\mathbf{D}}' = \mathbf{R}| \, \underline{\mathbf{D}} \mathbf{R}|^{\ddagger}, \quad \underline{\mathbf{A}}' = \mathbf{R}| \, \underline{\mathbf{A}} \mathbf{R}|^{\ddagger} \qquad (3.3.A)$$

where $\mathbf{R}|$ is $\mathbf{R}_S|$ for a spatial rotation and $\mathbf{R}_T|$ for a temporal rotation. Next we consider the transformation formulae,

$$\underline{\mathbf{M}}' = \mathbf{R}|^n \underline{\mathbf{M}} \mathbf{R}|^{\ddagger n}, \quad \underline{\mathbf{\Phi}}' = \mathbf{R}| \, \underline{\mathbf{\Phi}} \, \mathbf{R}|^{\ddagger n} \qquad (3.3.B)$$

With the understanding that a matrix to the power zero is the unit matrix, $n = 0$ gives the transformational properties suggested by Dirac [Dirac 1958]. However, $n = 1$ will also leave the Dirac equation invariant and here **M** and $\phi_1$ transform like four-vectors. In section 4 we show that the Dirac current will transform like a four-vector for this form of behaviour too. Since the transformational properties of **D** and **A** have not been altered, the eigenvalues associated with the trim Dirac equation for $n = 1$ continue the same as those of the original Dirac equation where $n = 0$.

As for the original, the trim Dirac equation is invariant under a parity change or time reversal. We have,

$$\underline{\mathbf{D}}'' = \underline{\mathbf{B}} \, \underline{\mathbf{D}} \, \underline{\mathbf{B}}^{\ddagger \mathrm{T}} \qquad (3.3.C)$$

and similarly for $\underline{\mathbf{A}}$, and,

$$\underline{\Phi}'' = \underline{\mathbf{B}}\,\underline{\Phi}\,\underline{\mathbf{E}}^{\ddagger\mathrm{T}}, \quad \underline{\mathbf{M}}'' = \underline{\mathbf{E}}\,\underline{\mathbf{M}}\,\underline{\mathbf{E}}^{\ddagger\mathrm{T}} \tag{3.3.D}$$

where,

$$\underline{\mathbf{B}} = \underline{\mathbf{C}}_{+}(1,1), \quad \underline{\mathbf{E}} = \underline{\mathbf{C}}_{-}(-1,1) \tag{3.3.E}$$

for a parity change, and,

$$\underline{\mathbf{B}} = \underline{\mathbf{C}}_{-}(-1,1), \quad \underline{\mathbf{E}} = \underline{\mathbf{C}}_{+}(1,1) \tag{3.3.F}$$

for time reversal. Either choice leaves equation (2.2.J) invariant. We leave open which circumstances permit complex conjugation, a reflection of the imaginary axes and thus terrestrial spacetime.

## 4. THE RADIATION EQUATION

### 4.1 The Dirac Current

The conserved current for the original Dirac equation is

$$J_0 = \Psi^{\dagger}\Psi, \quad J_r = \Psi^{\dagger}\alpha_r\Psi \tag{4.1.A}$$

Translating this into our new variables using equations (2.1.C) and (2.1.E),

$$J_0 = \left(\phi_1^{\dagger}\phi_1 + \phi_2^{\dagger}\phi_2\right)/2, \quad J_r = \left(\phi_2^{\dagger}\sigma_r\phi_2 - \phi_1^{\dagger}\sigma_r\phi_1\right)/2 \tag{4.1.B}$$

Changing variables using equations (1.3.C) and (2.1.L) for the first and (1.3.I) and (2.1.L) for the second,

$$\begin{aligned} J_0 &= \left(\phi_1^{\ddagger}\phi_1 + \phi_2^{\ddagger}\phi_2\right)/2 \\ J_r &= \left\{\left(\phi_1^{\ddagger}\mathbf{i}_r\phi_1 + \phi_2^{\ddagger}\mathbf{i}_r^{\ddagger}\phi_2\right)\mathbf{i}_3 + \mathbf{i}_3\left(\phi_1^{\ddagger}\mathbf{i}_r\phi_1 + \phi_2^{\ddagger}\mathbf{i}_r^{\ddagger}\phi_2\right)\right\}/4 \end{aligned} \tag{4.1.C}$$

We shall discuss a *Minkowski* version first in which,

$$J_0^{\sim} = iJ_0, \quad J_r^{\sim} = J_r \tag{4.1.D}$$

Pre- and post-multiplying by $1 + i\mathbf{i}_3$ ,

$$\begin{aligned} J_0^{\sim} &= t:\left\{i\left(1 + i\mathbf{i}_3\right)\left(\phi_1^{\ddagger}\phi_1 + \phi_2^{\ddagger}\phi_2\right)\left(1 + i\mathbf{i}_3\right)\right\}/4 \\ J_r^{\sim} &= t:\left\{i\left(1 + i\mathbf{i}_3\right)\left(\phi_1^{\ddagger}\mathbf{i}_r^{\ddagger}\phi_1 + \phi_2^{\ddagger}\mathbf{i}_r\phi_2\right)\left(1 + i\mathbf{i}_3\right)\right\}/4 \end{aligned} \tag{4.1.E}$$

where $t:(.)$ instructs us to take the temporal part. Setting,

$$\mathbf{J}_\mu^1 = i\mathbf{T}^+\phi_1^{\ddagger}\mathbf{i}_0\mathbf{i}_\mu^{\ddagger}\phi_1\mathbf{T}^-, \quad \mathbf{J}_\mu^2 = i\mathbf{T}^-\phi_2^{\ddagger}\mathbf{i}_0^{\ddagger}\mathbf{i}_\mu\phi_2\mathbf{T}^+, \tag{4.1.F}$$
$$\mathbf{T}^- = \mathbf{T}^+ = \frac{1 + i\mathbf{i}_3}{2}$$

we find that,

$$J_\mu^{\sim} = t:\left(\mathbf{J}_\mu^1 + \mathbf{J}_\mu^2\right) \tag{4.1.G}$$

Let,

$$\mathbf{J}_\mu^{\sim}| = \mathbf{J}_\mu^{\sim}|\left(\mathbf{J}_\mu^1, \mathbf{J}_\mu^2\right), \quad \underline{\mathbf{T}} = \underline{\mathbf{T}}\left(\mathbf{T}^+, \mathbf{T}^-\right), \quad \underline{\mathbf{I}}_\mu = \underline{\mathbf{I}}_\mu\left(\mathbf{i}_\mu, \mathbf{i}_\mu^{\ddagger}\right) \tag{4.1.H}$$

and we see from equations (4.1.F) that,

$$\mathbf{J}_\mu^{\sim}| = i\underline{\mathbf{T}}\,\underline{\mathbf{\Phi}}^{\ddagger\underline{\mathrm{T}}}\,\underline{\mathbf{I}}_0\underline{\mathbf{I}}_\mu\underline{\mathbf{\Phi}}\,\underline{\mathbf{T}} \tag{4.1.I}$$

giving,

$$J_\mu^{\sim} = t:\left\{\mathrm{Trace}\left(\mathbf{J}_\mu^{\sim}|\right)\right\} \tag{4.1.J}$$

from equation (4.1.G).

## 4.2 Invariances of the Dirac Current

We show that the current behaves like a Euclidean four-vector. We define the properties of $\underline{\mathbf{T}}$ and $\underline{\mathbf{I}}_\mu$ under spatial rotation and Lorentz transformation,

$$\underline{\mathbf{T}}' = \mathbf{R}|^{n}\,\underline{\mathbf{T}}\,\mathbf{R}|^{\ddagger n}, \quad \underline{\mathbf{I}}'_{\mu} = \mathbf{R}|\,\underline{\mathbf{I}}_{\mu}\mathbf{R}|^{\ddagger} \qquad (4.2.A)$$

where $\mathbf{R}|$ is $\mathbf{R}_S|$ for a spatial rotation and $\mathbf{R}_T|$ for a temporal rotation. Equation (4.1.I) becomes

$$\mathbf{J}_{\mu}^{\sim}|' = i\underline{\mathbf{T}}'\underline{\mathbf{\Phi}}'^{\ddagger\mathrm{T}}\underline{\mathbf{I}}'_0\,\underline{\mathbf{I}}'_{\mu}\underline{\mathbf{\Phi}}'\underline{\mathbf{T}}' \qquad (4.2.B)$$

and hence from equations (3.3.B) and (4.2.A) we have for rotations,

$$\mathbf{J}_{\mu}^{\sim}|' = \mathbf{R}|^{n}\mathbf{J}_{\mu}|\,\mathbf{R}|^{\ddagger n} \qquad (4.2.C)$$

where $\mathbf{R}|$ is $\mathbf{R}_S|$ for a spatial rotation and $\mathbf{R}_T|$ for a temporal rotation. Neither transformation alters the temporal part of the quaternion elements of $\mathbf{J}_{\mu}^{\sim}|$. Hence,

$$J_{\mu}^{\sim} = t:\left\{\mathrm{Trace}\left(\mathbf{J}_{\mu}^{\sim}|'\right)\right\} \qquad (4.2.D)$$

from equation (4.1.J) and so $J_{\mu}^{\sim}$ is invariant under spatial rotation and Lorentz transformation for both $n = 0$ and $n = 1$. We may then let,

$$\mathbf{J}^{\sim} = \sum_{\mu} J_{\mu}^{\sim}\mathbf{i}_{\mu} \qquad (4.2.E)$$

and we see immediately that $\mathbf{H}^{-1}\left(\mathbf{J}^{\sim}\right)$ is a four-vector.

We may also show that the current is conserved in all frames. Premultiplying equation (2.2.J) by $\underline{\mathbf{\Phi}}^{\ddagger\mathrm{T}}\underline{\mathbf{I}}_0$,

$$\underline{\mathbf{\Phi}}^{\ddagger\mathrm{T}}\underline{\mathbf{I}}_0\overrightarrow{\underline{\mathbf{D}}\underline{\mathbf{\Phi}}} = \underline{\mathbf{\Phi}}^{\ddagger\mathrm{T}}\underline{\mathbf{I}}_0\underline{\mathbf{\Phi}}\underline{\mathbf{M}} \qquad (4.2.F)$$

where the arrow indicates the direction of differentiation. Pre-multiplying equation (2.2.J) by $\underline{\mathbf{I}}_0$, taking the quaternion conjugate and reflector transpose, and post-multiplying the result by $\underline{\mathbf{\Phi}}$,

$$\overleftarrow{\underline{\mathbf{\Phi}}^{\ddagger\underline{\mathrm{T}}}\underline{\mathbf{I}}_0\underline{\mathbf{D}}}\underline{\mathbf{\Phi}}=\underline{\mathbf{M}}^{\ddagger\underline{\mathrm{T}}}\underline{\mathbf{\Phi}}^{\ddagger\underline{\mathrm{T}}}\underline{\mathbf{I}}_0\underline{\mathbf{\Phi}} \tag{4.2.G}$$

since,

$$\left(\overleftarrow{\underline{\mathbf{I}}_0\underline{\mathbf{D}}}\right)^{\ddagger\underline{\mathrm{T}}}=\overleftarrow{\underline{\mathbf{I}}_0\underline{\mathbf{D}}} \tag{4.2.H}$$

Adding equations (4.2.F) and (4.2G), multiplying by $i$, pre- and post-multiplying by $\underline{\mathbf{T}}$ and equating the traces,

$$\sum_{\mu}\left[\partial/\partial x_{\mu}\left\{t:\mathrm{Trace}\left(i\underline{\mathbf{T}}\underline{\mathbf{\Phi}}^{\ddagger\underline{\mathrm{T}}}\underline{\mathbf{I}}_0\underline{\mathbf{I}}_{\mu}\underline{\mathbf{\Phi}}\underline{\mathbf{T}}\right)\right\}\right]= \tag{4.2.I}$$
$$t:\mathrm{Trace}\left(i\underline{\mathbf{T}}\underline{\mathbf{\Phi}}^{\ddagger\underline{\mathrm{T}}}\underline{\mathbf{I}}_0\underline{\mathbf{\Phi}}\underline{\mathbf{M}}\underline{\mathbf{T}}+i\underline{\mathbf{T}}\underline{\mathbf{M}}^{\ddagger\underline{\mathrm{T}}}\underline{\mathbf{\Phi}}^{\ddagger\underline{\mathrm{T}}}\underline{\mathbf{I}}_0\underline{\mathbf{\Phi}}\underline{\mathbf{T}}\right)$$

Spatial or temporal rotations of the expressions in round brackets on both sides of equation (4.2.I) induce a similarity transformation, for both $n=1$ and $n=0$ transformational behaviour. The expression in round brackets on the right hand side of the equation is zero when $\mathbf{M}=m$, since

$$\underline{\mathbf{M}}=-\underline{\mathbf{M}}^{\ddagger\underline{\mathrm{T}}}, \tag{4.2.J}$$

and remains so in all frames. The expression in curly brackets on the left hand side of the equations equals the Minkowski current, $J_{\tilde{\mu}}$, from equations (4.1.I) and (4.1.J) and is constant from equation (4.2.D). Hence $J_{\tilde{\mu}}$ is conserved in all frames for all values of $n$.

In the above $D_{\mu}$ does not vary. If we wish to vary $D_{\mu}$ as a four-vector, we must induce a similar variation for $J_{\tilde{\mu}}$ by replacing it with $\mathbf{H}^{-1}(\mathbf{J})$. Then the left-hand side of equation (4.2.J) is equal to the dot product of four-vectors if $\mathbf{M}$ is replaced by $\mathbf{M}^{\sim}$ and $D_{\mu}$ is replaced by $iD_{\mu}$, and is hence constant and thus always zero.

## 4.3 A Simpler Version of the Dirac Current

This definition of $\mathbf{J}^{\sim}$ is not without problems. In the first place it would lead to a scalar temporal part for the potential, while we would like the possibility of making this triune as we suggested in sections 2.3 and 2.4. Secondly, it appears to work only if we use the Minkowski current defined in equation (4.1.D). Thirdly, the definition of $\mathbf{J}_{\mu}^{\sim}|$ in equation (4.1.I) is very complicated. We would prefer to choose an alternative. Since $(J_0, J_r)$ is a four-vector, there is a frame in which it has only one non-zero component, $J_0$. We see from equations (4.1.C) that in this frame $\phi_1 = \phi_2$, so that, the quaternion associated with the Dirac current,

$$\mathbf{J} = \phi_1^{\ddagger}\phi_1 \tag{4.3.A}$$

Then, in the same frame, let,

$$\mathbf{J}| = \mathbf{J}|\left(\mathbf{J}, \mathbf{J}^{\ddagger\underline{\mathrm{T}}}\right) \tag{4.3.B}$$

giving,

$$\mathbf{J}| = \underline{\mathbf{\Phi}}^{\ddagger\underline{\mathrm{T}}}\underline{\mathbf{\Phi}} \tag{4.3.C}$$

We make $\mathbf{H}^{-1}(\mathbf{J})$ a four-vector by amending the equation to read,

$$\underline{\mathbf{J}}\left(\mathbf{J}, \mathbf{J}^{\ddagger}\right) = \underline{\mathbf{\Phi}}^{\ddagger\underline{\mathrm{T}}}\underline{\mathbf{\Phi}}\,\underline{\mathbf{K}}\left(\mathbf{k}, \mathbf{k}^{\ddagger}\right) \tag{4.3.D}$$

where $\mathbf{k} = 1$ in the frame where $\phi_1 = \phi_2$ and we define the behaviour of $\underline{\mathbf{K}}$ under a spatial or temporal rotation as,

$$\underline{\mathbf{K}'} = \mathbf{R}|\,\underline{\mathbf{K}}\,\mathbf{R}|^{\ddagger} \tag{4.3.E}$$

We then induce a spatial or temporal transformation to the dashed frame by,

$$\underline{\mathbf{J}}' = \mathbf{R}| \underline{\boldsymbol{\Phi}}^{\ddagger\mathrm{T}} \underline{\boldsymbol{\Phi}}\underline{\mathbf{K}}\mathbf{R}|^{\ddagger} \tag{4.3.F}$$

where $\mathbf{R}|$ is $\mathbf{R}_S|$ for a spatial rotation and $\mathbf{R}_T|$ for a temporal rotation. This is successful because $\underline{\boldsymbol{\Phi}}^{\ddagger\mathrm{T}}\underline{\boldsymbol{\Phi}}$ is invariant; it is a rotator whose quaternion constituents have only a temporal part. Since $\mathbf{H}^{-1}(\mathbf{J})$ is equal to the Dirac current in one frame, and behaves as a four-vector, it must be equal to the Dirac current in all frames. Following our argument in the latter part of section 4.2, we find that equation (4.2.I) may be replaced by,

$$\sum_{\mu}\left[\partial/\partial x_{\mu}\left\{\mathrm{Trace}\left(\underline{\boldsymbol{\Phi}}^{\ddagger\mathrm{T}}\underline{\mathbf{I}}_0\underline{\mathbf{I}}_{\mu}\underline{\boldsymbol{\Phi}}\right)/2\right\}\right]= \tag{4.3.G}$$
$$\mathrm{Trace}\left(\underline{\boldsymbol{\Phi}}^{\ddagger\mathrm{T}}\underline{\mathbf{I}}_0\underline{\boldsymbol{\Phi}}\underline{\mathbf{M}}+\underline{\mathbf{M}}^{\ddagger\mathrm{T}}\underline{\boldsymbol{\Phi}}^{\ddagger\mathrm{T}}\underline{\mathbf{I}}_0\underline{\boldsymbol{\Phi}}\right)/2$$

where $\mu$ has only one value, zero. By explicit enumeration,

$$\mathbf{J} = \mathrm{Trace}\left(\underline{\boldsymbol{\Phi}}^{\ddagger\mathrm{T}}\underline{\mathbf{I}}_0\underline{\mathbf{I}}_{\mu}\underline{\boldsymbol{\Phi}}\right)/2 \tag{4.3.H}$$

when $\mu = 0$, from equation (4.3.D) with $k = 1$. Then, by a similar argument to that already given in section 4.2, the Dirac current is conserved.

## 4.4 The Trim Radiation Equation

The radiation equation is [Greiner and Reinhardt 1996a],

$$\Box A_{\mu} - \frac{\partial}{\partial x_{\mu}}\left\{-\frac{\partial A_0}{\partial x_0}+\sum_{r}\left(\frac{\partial A_r}{\partial x_r}\right)\right\} = J_{\mu}, \tag{4.4.A}$$

In the Lorentz gauge this can be written,

$$\underline{\mathbf{D}}^{\sim}\underline{\mathbf{D}}^{\sim}\underline{\mathbf{A}}^{\sim} = -\underline{\boldsymbol{\Phi}}^{\ddagger\mathrm{T}}\underline{\boldsymbol{\Phi}}\underline{\mathbf{K}}^{\sim},\quad \underline{\mathbf{K}}^{\sim} = \underline{\mathbf{K}}^{\sim}\left(\mathbf{k}^{\sim},\mathbf{k}^{\sim\ddagger}\right), \tag{4.4.B}$$
$$\mathbf{k}_{\mathrm{r}}^{\sim} = \mathbf{k}_r,\quad \mathbf{k}_0^{\sim} = i\mathbf{k}_0$$

from the definition of the d'Alembertian, $\Box$, or from Bell et al. [2000] and equations (4.3.D). This is invariant under a Minkowski transformation from the definition of $\underline{\mathbf{D}}^{\sim}$ in equation (2.1.N) and equations (3.3.A), (3.3.B) and (4.3.E).

To prove that equation (4.4.A) is invariant under a Euclidean transformation, we will consider the frame in which $\mathbf{A}$ has only a temporal part. Equation (4.4.A) becomes

$$\sum_r \frac{\partial}{\partial x_0} \frac{\partial A_r}{\partial x_r} = J_0, \tag{4.4.C}$$

Then we use the Euclidean transformation, obtaining in the general case,

$$\sum_\mu \frac{\partial^2 A_\mu}{\partial x_\mu^2} - \frac{\partial}{\partial x_\mu} \left\{ \sum_\mu \frac{\partial A_\mu}{\partial x_\mu} \right\} = J_\mu, \tag{4.4.D}$$

from which, or, by a similar argument from equation (4.4.B), *the trim radiation equation* in the Lorentz gauge is

$$\underline{\underline{\mathbf{D}}}\underline{\mathbf{D}}\,\underline{\mathbf{A}} = \underline{\mathbf{\Phi}}^{\ddagger\mathrm{T}}\underline{\mathbf{\Phi}}\underline{\mathbf{K}} \tag{4.4.E}$$

using equation (4.3.D).

The left-hand side of equation (4.4.E) is invariant under spatial or temporal rotation from equation (3.3.A). We may demonstrate invariance under parity change and time reversal by transforming $\underline{\mathbf{D}}$, $\underline{\mathbf{A}}$ and $\underline{\mathbf{K}}$ as in equation (3.3.C), while $\underline{\mathbf{\Phi}}$ transforms as in equation (3.3.D).

So far in this account of the radiation equation we have not allowed $\mathbf{A}_0$ the triune nature we previously supposed it might have. To do so, we must impose a triune nature on $\mathbf{J}_0$, the first component of $\mathbf{J}$. We write the trim radiation equation as,

$$\underline{\mathbf{D}}\underline{\mathbf{D}}\,\underline{\mathbf{A}} = \underline{\mathbf{J}} \qquad (4.4.F)$$

and insist on,

$$\mathbf{J} = \mathbf{H}\{(\mathbf{J}_0,\mathbf{J}_1,\mathbf{J}_2,\mathbf{J}_3)\}, \quad \underline{\mathbf{J}} = \underline{\mathbf{J}}(\mathbf{J},\mathbf{J}^{\ddagger}), \qquad (4.4.G)$$
$$|\mathbf{J}_0| = J_0, \quad |(\mathbf{J}_1,\mathbf{J}_2,\mathbf{J}_3)| = -|(J_1,J_2,J_3)|,$$

where we assign values to $\mathbf{J}_\mu$ in conformity with those assigned to the potential. A triune nature for charge as has been previously suggested by Rowlands and Cullerne [2001].

We may use the radiation equation, (4.4.F), to represent both the terrestrial and the celestial case. In the rest frame the terrestrial and celestial coincide and then our previous discussion at the end of section 3.2 follows on.

**ACKNOWLEDGEMENTS**

One of us (Bell) would like to acknowledge the assistance of E. A. E. Bell. She would like to dedicate this paper to Russell Bell, who is spinning with the universe.

**APPENDIX**

In the frame in which $\phi_1 = \phi_2$, discussed in section 4.3, the trim Dirac equation becomes

$$\left(\underline{\mathbf{D}} - \underline{\mathbf{A}} + \underline{\tilde{\mathbf{M}}}(\mathbf{M},-\mathbf{M}^{\ddagger})\right)\underline{\mathbf{\Phi}} = 0 \qquad (A.1)$$

provided $\mathbf{M} = m$, from equations (2.2.H) and (2.2.J). We call this *the special frame* and the equation *the special Dirac equation*. It is evident that this equation is relativistic too in both the terrestrial and celestial case, for the transformational behaviour given by equation (3.3.A) and,

$$\underline{\tilde{\mathbf{M}}'} = \mathbf{R}|\underline{\tilde{\mathbf{M}}}\mathbf{R}|^{\ddagger}, \qquad \underline{\boldsymbol{\Phi}'} = \mathbf{R}|\underline{\boldsymbol{\Phi}}\mathbf{R}|^{\ddagger n} \tag{A.2}$$

where $\mathbf{R}| = \mathbf{R}_S|$ for a spatial rotation and $\mathbf{R}| = \mathbf{R}_T|$ for a temporal rotation. The special Dirac equation, (1.A), has the same eigenvalues as the trim Dirac equation since $\underline{\mathbf{D}}$ and $\underline{\mathbf{A}}$ are unaltered.

We may use the special instead of the trim Dirac equation in section 2.4. Following through our argument there, we find that the Dirac equation with a source term, (2.4.B), is now covariant for four-vector behaviour of the bispinor, $\phi_1$.